\documentclass[referee]{aa-6-mpa}
\usepackage{graphicx}
\usepackage{natbib}
\usepackage{amssymb}
\usepackage{txfonts}
\usepackage{natbib}

\newcommand{\Iso}[2]{^{#1}{\rm #2}}
\newcommand{\msun}{M_\odot}
\newcommand{\lsun}{L_\odot}
\newcommand{\teff}{T_\mathrm{eff}}

\newcommand{\feh}{\mathrm{[Fe/H]}}

\newcommand{\amlt}{$\alpha_\mathrm{MLT}$}

\begin{document}

\title{$\alpha$-element enhanced opacity tables and low-mass metal-rich
stellar models\thanks{submtitted for publication to Astronomy \& Astrophysics}}

\author{A.~Weiss\inst{1} \and M.~Salaris\inst{2,1} \and
        J.W.~Ferguson\inst{3} \and D.R.~Alexander\inst{3}}

\institute{Max-Planck-Institut f\"ur Astrophysik,
           Karl-Schwarzschild-Str.~1, 85748 Garching,
           Federal Republic of Germany
           \and
           Astrophysics Research Institute, Liverpool John Moores University,
           Twelve Quays House, Egerton Wharf, Birkenhead, CH41 1LD, UK
           \and
           Physics Department, Wichita State University, Wichita,
           KS~67260-0032, USA
           }


\date{Received; accepted}

\authorrunning{Weiss et al.}
\titlerunning{$\alpha$-enhanced opacity tables}

\abstract 
{Calculation of stellar models with varying degrees of
  $\alpha$-element enhancement.}
{We investigate the influence of both a new generation of
  low-temperature opacities and of various amounts of  $\alpha$-element
  enhancements on stellar evolution models.}
{New stellar models with two different $\alpha$-element mixtures and
  two sets of appropriate opacity tables are computed and
  compared. The influence of the different mixtures as well as that of
  the improved generation of opacity tables is investigated.}
{It is found that around solar metallicity the new opacity tables have
  a drastic influence on stellar temperatures, which is mainly an
  effect of the new low-temperature tables, and not of variations in $\alpha$-element
  enhancement factors. The latter, however, influence stellar
  lifetimes via systematic opacity effects at core temperatures. We
  trace the reason for the low-temperature table changes to errors in
  the old tables.}
{Variations in $\alpha$-element abundance ratios affect the
  main-sequence properties  of super-solar metallicity stars
  significantly. Red giant branch effective temperatures depend only
  slightly on the specific mixture. Our older 
  low-temperature opacity tables were shown to be erroneous and should
  no longer be used for 
  stellar models with near- or super-solar metallicity. Corrected
  tables have already been produced.}


\maketitle

\clearpage

\section{Introduction}

It is now widely accepted that the chemical composition of stars can
be classified by two main classes of metal mixtures: one with nearly
solar metal ratios and another one, in which the so-called
$\alpha$-elements are enriched, while most other elements, in
particular C, N, and those of the iron-group, are present in
near-solar element ratios. The $\alpha$-elements are those produced by
$\alpha$-captures on oxygen in Supernova II explosions; 
they are O, Ne, Mg, Ca, Si, Ti. These elements are enhanced over their
solar relative number fractions by factors of 1.5-3 (or 0.2-0.5~dex on
a logarithmic scale). Since oxygen is contributing around 50\% of the
metals, its $\alpha$-enhancement can significantly increase the total
metallicity. Using one element, in particular iron, as the metal
indicator, can therefore be misleading in evaluating the total
metallicity of stars. In the traditional picture of galactic
populations, Pop~I is characterized by a solar metal mixture, while
Pop~II shows significant $\alpha$-enhancement. However, the
correlation with total metallicity or [Fe/H] is not at all a strict
one, since the relative $\alpha$-element abundances depend on the star
formation history and rate. It is known that in elliptical galaxies
and possibly the galactic bulge stellar compositions with solar or
super-solar iron abundances and strong $\alpha$-enhancement exist,
indicating intensive and short-lived star formation histories. For a
discussion of $\alpha$-element evolution and its connection with
galacto-chemical evolution, see the review by \citet{wst:89}.

Calculation of stellar models therefore requires for each value of
total metallicity ($Z$) at least these two basic internal metallicity
mixtures: a solar-scaled and an $\alpha$-enhanced one. This
requirement has been realized in the early nineties and since then the
number of stellar libraries which also take into account
$\alpha$-enhancement has steadily increased. While the inclusion of
individual element abundances is trivial for the nuclear energy
generation, it is non-trivial for an elaborate equation of state and for
opacities. In the former case, one trusts that as long as hydrogen and
helium are dominating the gas, the individual element abundances
within the metals is of negligible influence. However, for the
opacities this no longer holds true since the various elements contribute
very differently to the total absorption. As the calculation of
opacities has developed into a specialised branch, this requires
separate calculations by groups active in this field or flexible
programs for end-users.

In the past we have tried to include $\alpha$-enhancement in the
opacity tables we used as best as possible. In \citet{bct:89} an
$\alpha$-enriched mixture was first taken into account using the
Los~Alamos Astrophysical opacity library \citep{hmm:77} for
temperatures above 1~eV. Similar tables were used in \citet{wpm:95},
where for the first time $\alpha$-enhanced stellar models for
elliptical galaxies were calculated with almost consistent input
physics. However, at low temperature the opacities were still
solar-scaled. While it was shown by \citet{ssc:93} that at low
metallicities opacities with solar-scaled abundances and essentially
the same total metallicity are presumably a good replacement for
$\alpha$-enhanced tables, this could not be assumed for near-solar or
even super-solar metallicities. The final step became possible when
low-temperature opacity tables, which include molecular absorption
processe,s were made available by \citet[AF94]{af:94}, both for solar and
$\alpha$-enhanced metal mixtures. These could then be combined with
high-temperature tables for {\em completely identical} mixtures from
\cite[OPAL; IR96]{ir:96}. The enhancement for the various $\alpha$-elements
are variable in this metal mixture \citep{sw:98}.
These tables were used by the authors for globular
cluster calculations \citep{sw:98} demonstrating also that the
approximation by \citet{ssc:93} begins to fail around $\feh=-0.7$,
the iron content of 47~Tuc. A complete set of models with
metallicities ranging from $Z=0.008$ to $0.070$ using the same opacity
tables was added to the Padova stellar library by \citet{sgwc:2000}. Similar, but
independent opacity tables from the same sources were used, e.g., in
the models by the Victoria and Yale-Yongsei groups
\citep{vsria:2000,ykd:2003}. Here, all $\alpha$-elements have been
enhanced by the same factor, ranging in the different sets from 0.3 to
0.6~dex. 

As demonstrated first by \citet{ssc:93}, and verified by, e.g.,
 \citet{sw:98} and \citet{vsria:2000},  to first order
$\alpha$-enhanced opacities at low metallicity and for low mass
stellar models can be replaced by scaled-solar
ones for the same metal fraction $Z$, provided that the ratio
$(X_\mathrm{C}+X_\mathrm{N}+X_\mathrm{O}+X_\mathrm{Ne})/(X_\mathrm{Mg}+X_\mathrm{Si}
 +X_\mathrm{S}+X_\mathrm{Ca}+X_\mathrm{Fe})$  -- where
$X_\mathrm{i}$ denotes the mass fraction of element i -- is approximately
the same in both mixtures (in spectroscopic notation this corresponds to
[$(X_\mathrm{C}+X_\mathrm{N}+X_\mathrm{O}+X_\mathrm{Ne})/(X_\mathrm{Mg}+
 X_\mathrm{Si}+X_\mathrm{S}+X_\mathrm{Ca}+X_\mathrm{Fe})$]$\sim 0$).
As a consequence, differences in the internal distribution of
$\alpha$-elements are of minor significance, as long as this condition
is satisfied.
The extension of this result to solar or super-solar metallicities
has however never been investigated.

For a population
synthesis project (Coelho et al.~2006, in preparation), we decided to
compare stellar models of low-mass stars using the previously
introduced $\alpha$-enhanced tables with varying enhancement factors
to other models with new opacity tables, in which all
$\alpha$-elements are enhanced by a constant factor of 0.4~dex. 
The individual metal
abundances are listed in Table~\ref{t:1} and provide
$[(X_\mathrm{C}+X_\mathrm{N}+X_\mathrm{O}+X_\mathrm{Ne})/(X_\mathrm{Mg}+X_\mathrm{Si}
 +X_\mathrm{S}+X_\mathrm{Ca}+X_\mathrm{Fe})]=0.14$
in case of the variable enhancement mixture, and 0.007 for the constant
enhancement one. 
The
total metallicity ranges from $Z=0.011$ to $0.048$. For the
calculation of the latter tables the new low-temperature molecular
opacity code by \citet[FA05]{fa:2005} and for the high-temperature
regime the OPAL online calculation of opacity tables\footnote{\tt
http://www-phys.llnl.gov/Research/OPAL} were used. We therefore are
faced with the fact that the new tables (for $\log T \leq 4.5$; but
used only for $\log T \leq 4.1$) differ from the old ones both in the
internal metal distribution (``mixture effect'') and in the method of
calculation (``generation effect''). For the higher temperatures only
the mixture effect is present.

The purpose of the present paper is to present new
low-mass stellar models up to the red giant tip and the influence of
variations in $\alpha$-enhancement on the evolution at given total
super-solar metallicity. Second,
we investigate the influence of the new low-temperature opacities and
that of a variation of internal element distributions in these
tables. This is an important issue because of the fact that
one cannot expect to have always opacities at hand which are fully
consistent with the mixture chosen for the stellar models.
For completeness, we also discuss metal-poor cases.

The stellar models and the sets of opacity tables are
introduced in Sect.~2. The basic results of the calculations and
various tests are the subject of Sect.~3.
The low-metallicity case follows in Sect.~5, before the
discussion and conclusions close the paper in the final section.

\section{Model details}

\subsection{Stellar evolution program}\label{s:prog}

The stellar models were calculated with the Garching stellar evolution
code, which was described in \citet{wsch:2000} and lately in
\citet{wsksc:2005}. We will concentrate here on the description of the
opacity tables. Our code uses tables of Rosseland mean opacities
$\kappa$ for
mixtures quantified by the mass fractions of hydrogen and metals (both
ranging from 0 to 1; in total of order 80), and with a temperature and
density grid of about 85 and 25 grid points\footnote{As the density
coordinate we actually use the usual $\log R = \log
\rho-3\log T+18$}. The interpolation in this grid is done by a
two-dimensional, bi-rational spline algorithm
\citep{spaeth:73}. In mixture we use parabolic polynomials first in
$X$ (hydrogen) between the three tables closest to the actual value,
and then in $\log Z$ (metallicity). In practice, and as long as the
total metallicity is not changing, e.g.\ either by advanced nuclear
burning phases or metal diffusion, tables for only three metallicity
values are sufficient for the whole calculation. For evolutionary
stages from core helium burning on, we have special core tables, but
these are not of interest for the present work.

The tables themselves are the end product of the merging of source
data. We have four main regions in the $T-R$ domain:
\begin{itemize}  
\item $\log T < 3.8$ , for which we use the Wichita State Alexander \&
Ferguson molecular opacity tables mentioned in the introduction
\item $8.7 >\log T > 4.1$, for which OPAL tables are used
\item high density: here we employ the results by \citet{imi:83} for
electron conduction opacities
\item $\log T > 8.7$, for which no OPAL data are available; here we
use the old Los Alamos Opacity Library (see Sect.~1; in the present
case this is irrelevant).
\end{itemize}

In between these regions are transitions. Between $3.8 < \log T <
4.1$, where both Wichita State and OPAL data are available, we have a linear
transition in $\log \kappa$ along with $\log T$ from one source to the
other. The agreement between both tables is excellent and the
transition is very smooth. At the high-density edge we add radiative
($\kappa_\mathrm{r}$) and conductive ($\kappa_\mathrm{c}$) opacities
according to  
\[ 1/\kappa = 1/\kappa_\mathrm{r} + 1/\kappa_\mathrm{c}. \]
Since with increasing density $\kappa_\mathrm{c}$ is dominating, the
radiative contribution can be omitted once $\kappa_\mathrm{r} >
\kappa_\mathrm{c}$. However, in particular at $\log T < 5$ the
radiative tables end before this situation is reached. In case the gap
between the end of 
the radiative opacity table and the density from which on
$\kappa_\mathrm{c}$ is already lower than the last radiative value
available is only 1-2~dex, we boldly interpolate over this gap (cubic
spline). If the gap is too large, the final $\kappa$-table has to end
here. Should we run out of the table definition range during the
stellar model calculations, we use the last
table value. This happens at isolated points in some low-mass
main-sequence envelopes and cannot be avoided as long as the input
tables do not cover the full $\rho-T$-plane.

All our tables are constructed in this same way in a separate step 
before they are used by the stellar evolution code. 

\subsection{Stellar models}

For the larger project about stellar population synthesis (Coelho et 
al.\ 2006, in preparation) we calculated mixtures of varying metallicity from 
$Z=0.005$ to 0.048, but in the present paper we will be concentrating on one
mixture, which is $X=0.679$, $Z=0.032$. All $\alpha$-elements are
enhanced over the solar-scaled abundances by 0.4~dex, for which we
used the solar mixture by \citet{gs:98}. For comparison with a
solar-scaled mixture with identical [Fe/H] models with the mixture
$X=0.735$, $Z=0.017$ was chosen; the higher total metallicity is thus
assumed to be connected with an increase in helium abundance. In mass
our models cover the range from $0.6$ to $1.2\,\msun$ in steps of
$0.05\,\msun$. Our calculations start with the zero-age
main-sequence (ZAMS) defined by homogeneous stellar composition and
vanishing gravo-thermal energies in the initial model. Since this
composition is not identical to nuclear equilibrium in the core
($\Iso{3}{He}$, CNO-isotopes) the initial few million years are spent
to adjust the composition accordingly. This leads to a well-known
short loop around the ZAMS position, which will be visible in some of
the diagrams. We follow the evolution into the core helium flash at
the tip of the RGB, until the helium luminosity is of order
$1000\,\lsun$. 

We emphasize that the chemical composition in the models is always the
same ($\alpha$-c), as given in Table~\ref{t:1}. What is varied in the
calculations presented below, is just the internal composition of the
opacity tables.

\begin{table}
\caption{Chemical abundances of the metals. Columns 2 and 3 contain the
standard \citet{gs:98} solar abundances on a logarithmic scale where
the hydrogen abundance is 12.00 and as relative mass fractions. 
Columns 3 and 4 show the abundances of the $\alpha$-enhanced mixture
by \citet[``$\alpha$-v'']{sw:98} and columns 5 and 6 for a constant
$\alpha$-enhancement of 0.4~dex (``$\alpha$-c''). $\alpha$-elements
are printed in italics.} 
\label{t:1}
\begin{center}	
\begin{tabular}{l|rr|rr|rr}
\hline\noalign{\smallskip} \hline
el. & \multicolumn{2}{c|}{solar} &  \multicolumn{2}{c|}{$\alpha$-v} &
\multicolumn{2}{c}{$\alpha$-c} \\
\hline
      C & 8.52 & 0.173344 & 8.52 & 0.076535 & 8.52 & 0.083953 \\
      N & 7.92 & 0.053187 & 7.92 & 0.023483 & 7.92 & 0.024600 \\
{\it  O}& 8.83 & 0.482487 & 9.33 & 0.673656 & 9.23 & 0.573606 \\
{\it Ne}& 8.08 & 0.096446 & 8.37 & 0.083031 & 8.48 & 0.128651 \\
     Na & 6.33 & 0.001999 & 6.33 & 0.000883 & 6.33 & 0.001038 \\
{\it Mg}& 7.58 & 0.037597 & 7.98 & 0.041697 & 7.98 & 0.049009 \\
     Al & 6.47 & 0.003599 & 6.47 & 0.001589 & 6.47 & 0.001868 \\
{\it Si}& 7.55 & 0.040543 & 7.85 & 0.035717 & 7.95 & 0.052850 \\
      P & 5.45 & 0.000355 & 5.45 & 0.000157 & 5.45 & 0.000184 \\
{\it  S}& 7.33 & 0.021158 & 7.66 & 0.019972 & 7.73 & 0.036357 \\
     Cl & 5.50 & 0.000456 & 5.50 & 0.000201 & 5.50 & 0.000237 \\
     Ar & 6.40 & 0.005380 & 6.40 & 0.002375 & 6.40 & 0.002118 \\
      K & 5.12 & 0.000210 & 5.12 & 0.000093 & 5.12 & 0.000109 \\
{\it Ca}& 6.36 & 0.003735 & 6.86 & 0.005215 & 6.76 & 0.004869 \\
{\it Ti}& 5.02 & 0.000204 & 5.65 & 0.000384 & 5.42 & 0.000266 \\
     Cr & 5.67 & 0.000989 & 5.67 & 0.000437 & 5.67 & 0.000513 \\
     Mn & 5.39 & 0.000549 & 5.39 & 0.000242 & 5.39 & 0.000285 \\
     Fe & 7.50 & 0.073517 & 7.50 & 0.032459 & 7.50 & 0.037284 \\
     Ni & 6.25 & 0.004246 & 6.25 & 0.001874 & 6.25 & 0.002203 \\
\end{tabular}
\end{center}
\end{table}

\subsection{Opacity tables}

The first set of models (A) was calculated with the same opacity
tables as those used by \citet{sw:98} and \citet{sgwc:2000}, which are
a combination of OPAL and Wichita State opacities for the $\alpha$-v
mixture\footnote{These tables were originally provided by F.~Rogers
and D.~Alexander by private communication (1994/1995), but later made public on
the respective websites,  {\tt http://www-phys.llnl.gov/Research/OPAL}
and  {\tt http://webs.wichita.edu/physics/opacity/}.}
They were calculated at the time when the new Wichita State low-temperature
molecular opacities \citep{fa:2005} were not yet available, in
particular not for non-solar metal compositions, and under the
assumption that the slightly different table composition would not
influence the model properties significantly, such that this internal
inconsistency could be tolerated.

Set B followed after low-T tables for mixture $\alpha$-c had been
calculated. High-T OPAL tables were obtained from the OPAL-website for
the identical mixture. This is the most recent set of opacity tables
and fully consistent with the model compositions. The models for the
population synthesis project (Coelho et al.\ 2006, in preparation)
have been computed with it.

To analyse the results of the following section, additional sets of
tables were produced for various test cases. They consist of {\em
inconsistent} combinations of low- and high-T tables and are listed in
Table~\ref{t:2} as well. The final case R is a repetition of case A,
but the tables were recomputed with the same code by \citet{fa:2005}.

\begin{table}
\caption{Details about the tables of Rosseland mean opacities for the
various sets of stellar models. Low- and high-temperature ranges are
as described in Sect.~\ref{s:prog}. Mixture labels are as in
Table~\ref{t:1} and ``code'' entries refer to the relevant
publication, where the calculation method and details are
described. ``AF05'' refers to a re-computation in 2005 using the
original AF94 program.}
\label{t:2} 
\begin{center}	
\begin{tabular}{l|ll|ll}
\hline\noalign{\smallskip} \hline
    & \multicolumn{2}{c|}{low-T} &  \multicolumn{2}{c}{high-T} \\
Set &  mix & code & mix & code  \\
\hline
A & $\alpha$-v & AF94 & $\alpha$-v & IR96 \\
B & $\alpha$-c & FA05 & $\alpha$-c & IR96 \\
T1& $\alpha$-c & FA05 & $\alpha$-v & IR96 \\
T2& $\alpha$-v & AF94 & $\alpha$-c & IR96 \\
T3& $\alpha$-v & FA05 & $\alpha$-c & IR96 \\
R & $\alpha$-v & AF05 & $\alpha$-v & IR96 \\
\end{tabular}
\end{center}
\end{table}

\section{Results}\label{s:res}

\subsection{From old to new (Case A to B)}

\begin{figure}	
\begin{center}
\includegraphics[scale=0.60]{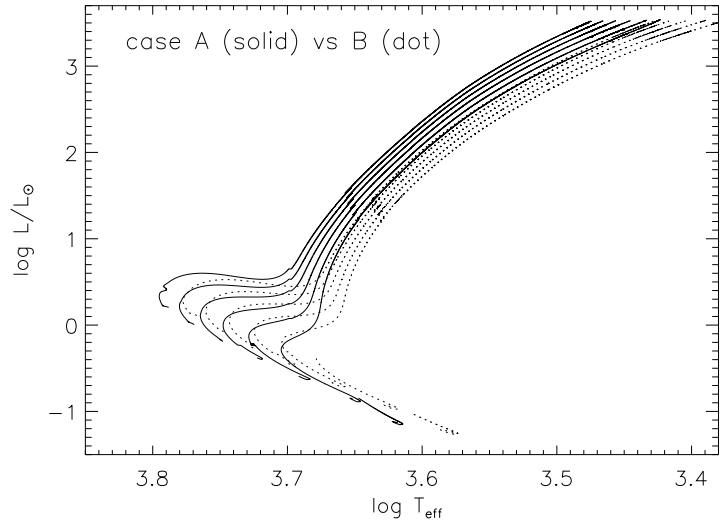}
\caption{Comparison of evolutionary tracks for cases A and B. Masses
range from $0.6\,\msun$ to $1.2\,\msun$ in steps of $0.1\,\msun$ (we
omitted the intermediate mass values). The
calculations were stopped either at the RGB tip or if the age exceeded
50~Gyr. For this reason the tracks do not extend to the RGB for
$M=0.60\,\msun$ for case A, and $M=0.60\,\msun$ and $0.70\,\msun$ for
case B. The composition of both model sets is identical ($\alpha$-c),
but that of the opacity tables is not.}
\label{f:1}
\end{center}
\end{figure}

The scientifically interesting question is how much the use of opacity
tables with an internal $\alpha$-element distribution ($\alpha$-v)
differing from that of the models themselves ($\alpha$-c) influences
the evolution. This is an interesting aspect of stellar modelling
as long as hghly accurate opacities cannot be calculated on-line with
the models. To evaluate this, we computed cases A and B, the latter
being the fully consistent one. We emphasize again that the model
composition in case A is the same as in B, but that the internal
metal distribution of the opacity tables is slightly different.
The resulting tracks are shown in
Fig.~\ref{f:1}. The effect is drastic: for the $0.9\,\msun$ model,
ZAMS and TO-positions are 0.08 resp.\ 0.06~dex brighter, the turn-off (TO) also
being 0.02~dex cooler and the main-sequence lifetime is 3.3~Gyr
shorter than that for case B (17.8~Gyr). RGB temperatures are higher
by $\approx 250$~K, too. Theses changes are also typical for all other
masses. 

In the tracks, small kinks are visible both on the main sequence and
on the subgiant branch. The reason is that in the phase lying between
these points the OPAL equation of
state \citep{rsi:96} does not cover the conditions in the outer
stellar envelope and we switch back to a simple Saha-type equation of
state. This effect is less pronounced for
lower stellar masses. We have repeated a few cases using the new
OPAL-based equation 
of state by Irwin (described in \citealt{csirwin:2003}), which is more
extended. The tracks from this repeated calculation agree extremely
well with those shown, but in addition are
smooth without any kinks. 

Since the strong influence of the different $\alpha$-enriched opacity
tables could be due to either the generation or the composition
effect, or both, we set out to isolate this by additional tests in which
we used specific opacity tables with only one effect being present.

\subsection{T1: new low-T opacities}\label{s:t1}

\begin{figure}	
\begin{center}
\includegraphics[scale=0.60]{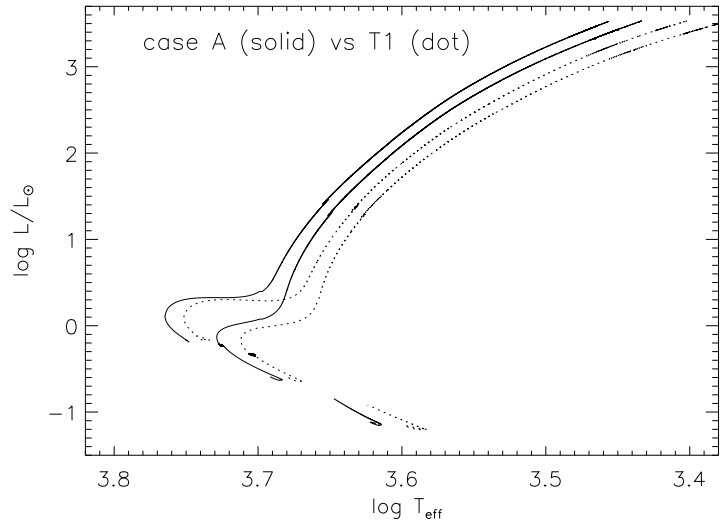}
\caption{Comparison of evolutionary tracks for cases A and T1. Masses
are $0.6,\,0.8$ and $1.0\,\msun$. }
\label{f:2}
\end{center}

\end{figure}

For this test case we used only the low-T opacity tables for the
$\alpha$-c mixture. The
high-T tables are still those for mixture $\alpha$-v as in case A.
Figure~\ref{f:2} shows the resulting tracks for three different
masses. Temperatures along the RGB (for $M=0.8$ and $1.0\,\msun$) are
again lower by more than 200~K for the new low-T tables, while the
ZAMS and TO luminosities differ only by a few hundreths of a
magnitude. Correspondingly, main-sequence lifetimes agree within
0.4~Gyr ($0.8\,\msun$ and $1.0\,\msun$). The new low-T opacities
apparently affect the envelope structure significantly, while the
high-T tables with the new mixture do influence the core properties. 

\subsection{T2: new high-T mixture}\label{s:t2}

\begin{figure}	
\begin{center}
\includegraphics[scale=0.60]{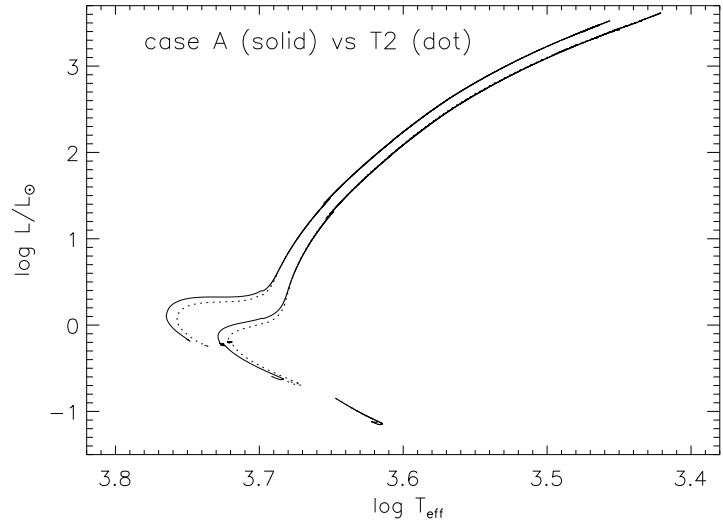}
\caption{Comparison of evolutionary tracks for cases A and T2. Masses
are $0.6,\,0.8$ and $1.0\,\msun$. }
\label{f:3}
\end{center}
\end{figure}

\begin{figure}	
\begin{center}
\includegraphics[scale=0.60]{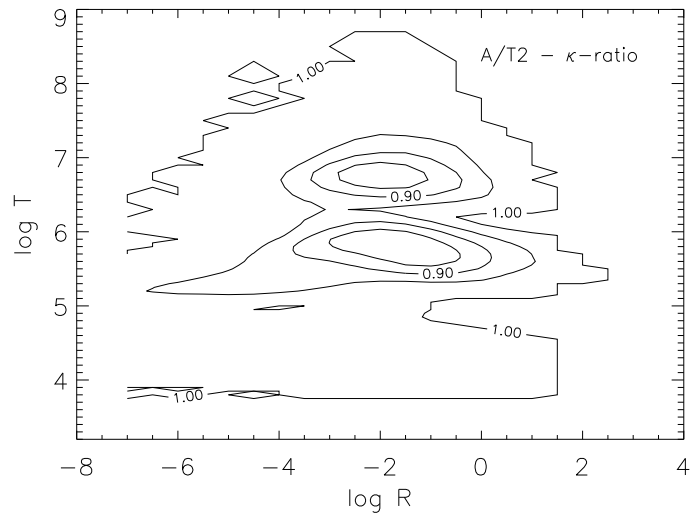}
\caption{Ratio of Rosseland opacities $\kappa$ between a typical table
used in cases A and T2. The composition is $X=0.700$, $Z=0.03$. Shown
are isocontours of the ratio.}
\label{f:4}
\end{center}
\end{figure}

Since the method for calculating the OPAL opacities has not changed,
replacing the high-T tables with those for the new mixture does not
imply a generation effect. Again, we show the comparison between case
A and this test case (T2) in Fig.~\ref{f:3}. Since the low-T opacities
are now identical, the RGB temperatures completely agree. However,
changes on the main-sequence remain. The ZAMS luminosity is lower by
0.07~dex in case T2 ($M=0.80\,\msun$), and the MS lifetime longer by
4.3~Gyr. The reason for this effect is evident from Fig.~\ref{f:4},
where we show a contour plot of the ratio of Rosseland opacities
between the old and new composition. Over a wide temperature and
density range, typical for core temperatures of main sequence stars
the $\alpha$-c opacities are higher by 5-20\%. The differences are
very similar also for lower helium content. An increase in
opacity, as is well known, leads to a lower luminosity and thus a
longer burning time. 

\subsection{T3: new generation of low-T tables}\label{s:t3}

\begin{figure}	
\begin{center}
\includegraphics[scale=0.60]{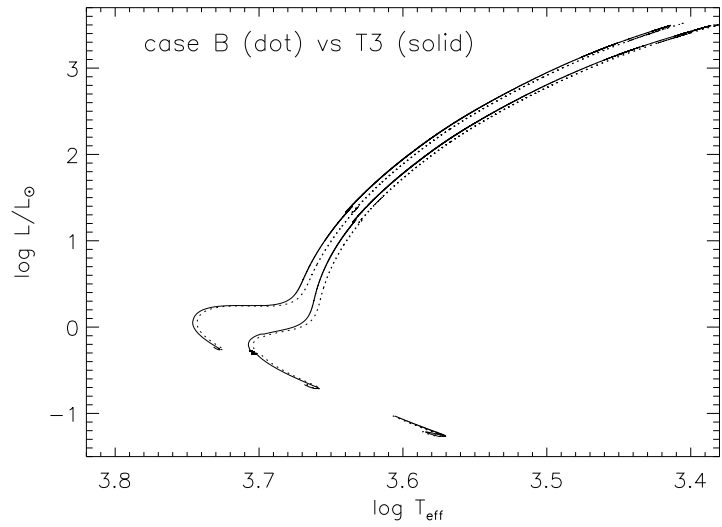}
\caption{Comparison of evolutionary tracks for cases B (dotted lines)
and T3 (solid) for masses of 
0.6, 0.8, and $1.0\,\msun$.}
\label{f:5}
\end{center}
\end{figure}

After we identified the main influence of low- and high-T tables, when
going from case A to B, we are separating generation- from mixture
effect for the low-temperature opacities. To this end, tables for
mixture $\alpha$-v were calculated using the method of \citet{fa:2005}
and combined with the $\alpha$-c OPAL tables. We thus compare to case
B, with only the mixture-effect at low temperatures present, shown in 
Fig.~\ref{f:5}. Obviously, in spite of
employing mixture $\alpha$-v for the molecular opacities, the track
resembles that of case B, the completely consistent $\alpha$-c
calculation, very much. In fact, for $M=0.80$ and $1.0\,\msun$ 
ZAMS luminosities agree within 0.003~dex, TO luminosities and
effective temperatures better than 0.004 resp.\ 0.003~dex, and RGB
tip temperatures better than 0.09~dex. The TO and RGB tip ages are
within 100~Myr. The conclusion from this experiment and the one of
Sect.~\ref{s:t1} is therefore that for the low-temperature opacities
the generation effect is the dominating one and that within the same
generation of tables the mixture effect is almost negligible. This is
different from the effect on the atomic opacities of the interior
(Sect.~\ref{s:t2}). 

\newpage

\subsection{R: low-T tables recalculated}\label{s:r}

\begin{figure}	
\begin{center}
\includegraphics[scale=0.60]{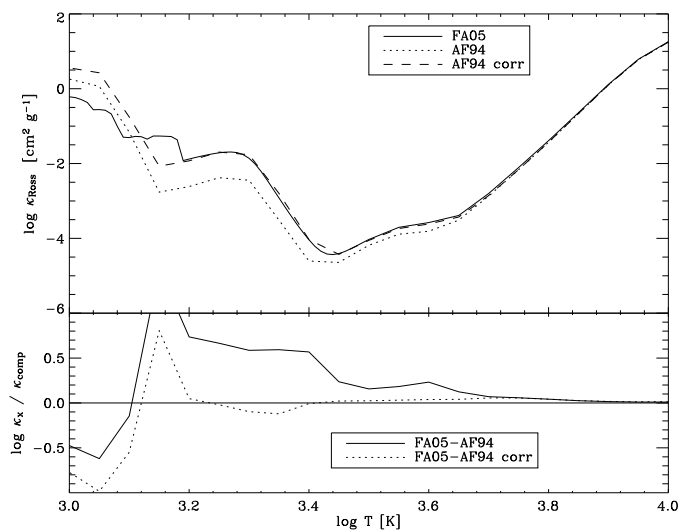}
\caption{Comparison of Rosseland mean opacities in a table with
mixture $X=0.7$, $Z=0.02$, taken at  $\log R = -3$ as function of
temperature (upper panel). Lower panel: differences in $\log
\kappa$. F05 denotes the new calculation of tables for the same
$\alpha-v$ element ratios as in the original AF94 table.}
\label{f:6}
\end{center}
\end{figure}

\begin{figure}
\begin{center}
\includegraphics[scale=0.60]{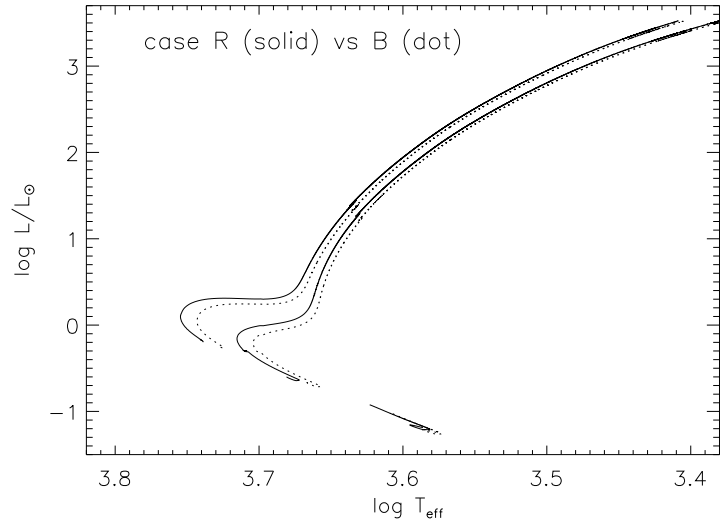}
\caption{Comparison of evolutionary tracks for cases B and R. Masses
are $0.6,\,0.8$ and $1.0\,\msun$.}
\label{f:7}
\end{center}
\end{figure}

The result of Sect.~\ref{s:t3} is quite surprising given the fact that
\citet{fa:2005} reported rather small changes in the new molecular
opacities for solar-type mixtures. To verify this conclusion further,
we investigated the low-temperature tables for $\alpha$-v mixtures  
in more detail. A comparison displayed quite large differences of $\pm
10$\% around $\log T \approx 3.6$, increasing further to lower
temperatures. Fig.~\ref{f:6} shows a sample comparison at $\log R =
-3$ for the mixture $X=0.7$, $Z=0.02$  between the old AF94 and the new
FA05 table. For $\log T \lesssim 3.4$ the difference in fact increases
up to a factor of 3. This finding emphasizes that the new calculation 
programs alone are responsible for the change in opacities. Using
these recalculated FA05 tables for mixture $\alpha$-v, combined with
the corresponding high-T OPAL tables, we show the comparison of this
case R, which is the ``2005 update'' of A, with case
B in Fig.~\ref{f:7}. We thus see here the effect of changing the 
$\alpha$-element enhancement factors within the opacity tables, which,
however, were calculated with the same program in both cases.
Obviously, along the RGB the two cases
agree very closely, while the differences in the earlier evolutionary
phases is that of Sect.~\ref{s:t2} and Fig.~\ref{f:3}.
						      
\subsection{$\alpha$-enhanced low-T tables of 1994}

In contrast to the solar metallicity case investigated in
\citet{fa:2005}, the re-calculation of the $\alpha$-v low-temperature
tables with the new molecular opacity code revealed the large changes
of Fig.~\ref{f:6}, in particular at $T \lesssim 2500$~K. The original
calculations by D.~Alexander (1994, private communication to A.~Weiss)
were therefore repeated with the original code of \cite{af:94}. The
result is also displayed in Fig.~\ref{f:6} as the dashed line, and
differs strongly from the original 1994 result, shown as the dotted
line, in the temperature region between $3.25 \lesssim \log T \lesssim
3.35$, while it agrees very well with the result from the new code, at
least down to $\log T \approx 3.15$, below which the differences can
be explained by by updated grain physics,
e.g.\ more grain species in the equation of state and updated grain
opacities, in FA05.

The conclusion is that an error was made in the original calculations
provided to A.~Weiss in 1994. This error was essentially a series of typos
in the input abundances to the opacity tables run at that time.
{\sl The $\alpha$-enhanced
tables used in several publications \citep[and
others]{sw:98,wsch:2000,sgwc:2000} are therefore erroneous and should
not be used any longer}. At least at solar and super-solar
metallicities they influence the effective temperatures of the models
strongly by the underestimation of the opacity at low
temperatures: Red Giant models are rendered too blue. Similar
opacity tables for different $\alpha$-enhancement factors calculated
with the same code and used, e.g.\ in \citet{vsria:2000} and
\citet{ykd:2003}, do not suffer from this error. 
We also emphasize that most of the papers using the erroneous tables
dealt with significantly lower total metallicity at which
the models are much less sensitive (see below).

\clearpage

\section{Low-metallicity models}

\begin{figure}[ht]
\begin{center}
\includegraphics[scale=0.60]{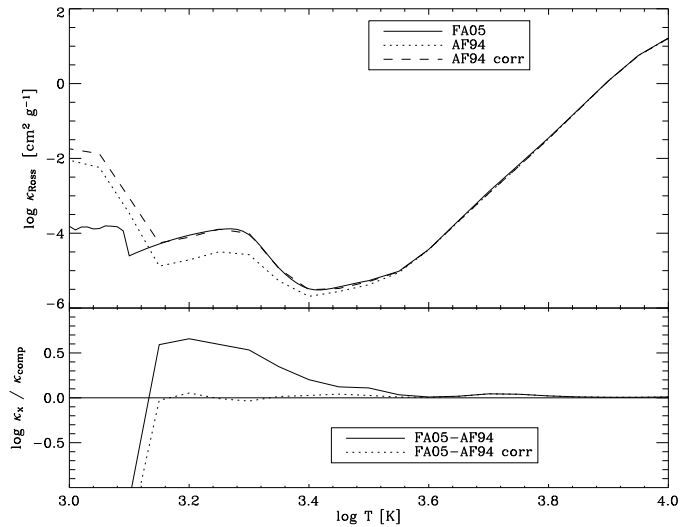}
\caption{As Fig.~\ref{f:6}, but for $Z=0.0001$.}
\label{f:8}
\end{center}
\end{figure}

Figure~\ref{f:8} shows the same opacity comparison between $\alpha$-c (``f05'')
and $\alpha$-v tables, the latter for the new (``af94 corr'') and
original (``af94 old'') calculation, for $Z=0.0001$. The situation resembles that of
Fig.~\ref{f:6}, but the opacity itself is lower by an order of
magnitude. All changes should therefore result in milder effects on
the stellar models. 

\begin{figure}	
\begin{center}
\includegraphics[scale=0.65]{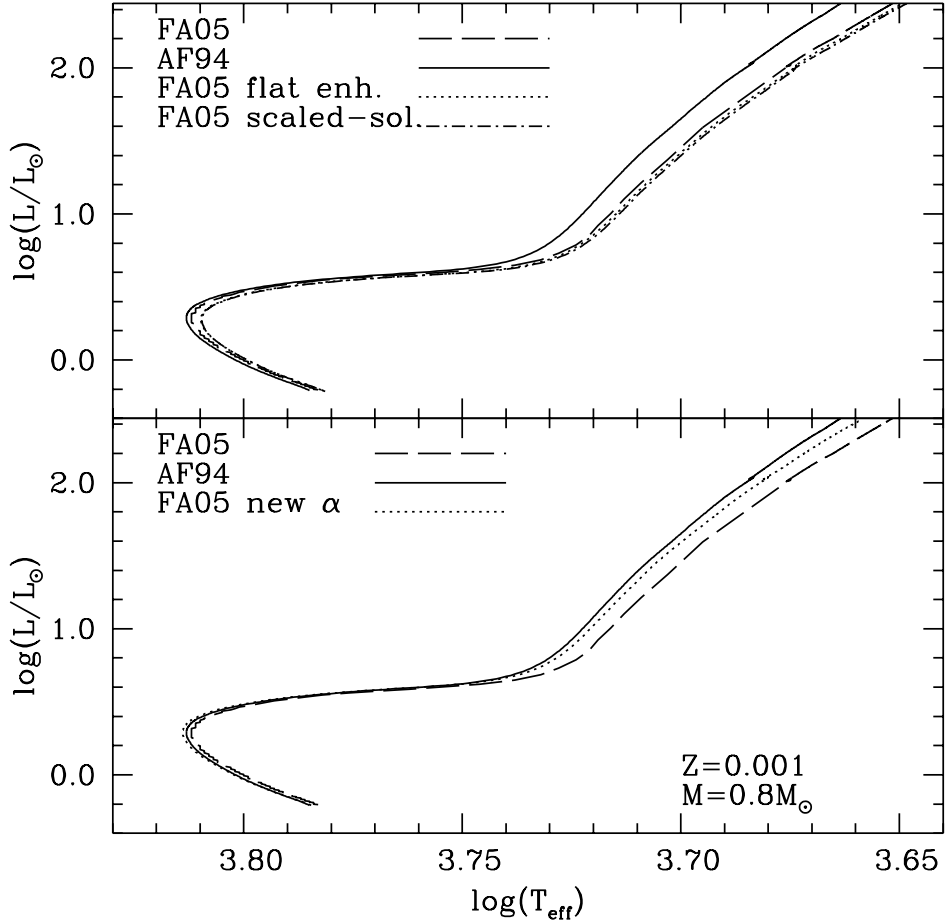}
\caption{Lower panel: Evolutionary tracks for an $M=0.8\,\msun$ star with $Z=0.001$
calculated with the original \citet[AF04]{af:94} and the new
\citet[F05]{fa:2005} code for the $\alpha$-v mixture. The dotted line
refers to a track with the new opacities, but a recalibrated mixing
parameter $\alpha_\mathrm{MLT}$. Upper panel: Effect of using opacity
tables with different internal element distributions but identical
$Z$; see text.}
\label{f:9}
\end{center}
\end{figure}

Indeed, Fig.~\ref{f:9} (lower panel) demonstrates this for the case of
$M=0.8\,\msun$, and $Z=0.001$: The track labelled ``AF94''
was calculated with the original $\alpha$-enhanced table and
corresponds to the tracks used in, e.g.\ \citet{sw:98}. The same code
as in this paper was used, too. Employing the
newly calculated opacities (F05) again results in a cooler RGB; all
other properties remain virtually unchanged. However, the difference in
$\teff$ amounts to less than 150~K compared to about 250~K for the
$Z=0.032$ cases discussed so far. 

\begin{figure}	
\begin{center}
\includegraphics[scale=0.65]{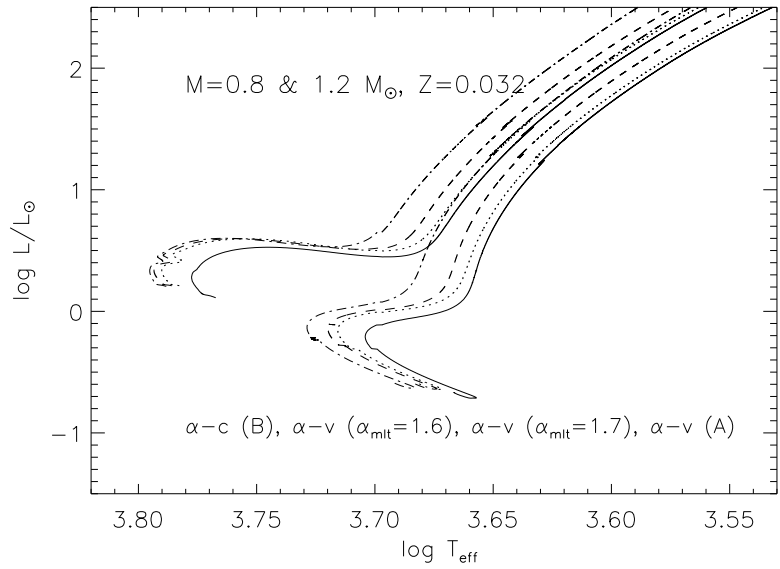}
\caption{Overview over various cases for two mass values (0.8 and
$1.2\,\msun$) for our standard composition of $X=0.679$, $Z=0.032$.
See text for more explanations.}
\label{f:11}
\end{center}
\end{figure}

The cooler RGB temperatures imply that the fits to real globular
cluster data will have 
to be reconsidered (they were, in fact, fitting quite well; see
\citealt{sw:97,sw:98}), a deeper look into this is warranted. With the
new generation of low-T tables also the opacities for solar mixtures
changed slightly \citep{fa:2005}, which implies that the solar models
need to be recalibrated and the parameter $\alpha_\mathrm{MLT}$ in the
mixing length theory needs some adjustment. Using the Garching
stellar evolution code A.~Serenelli (private communication, 2006)
found that $\alpha_\mathrm{MLT}$  had to be increased from 1.78 to
1.88. Similarly, for the code used for the models of Fig.~\ref{f:9},
a solar calibration (without diffusion) resulted in an increase by
0.12 to the value of 
$\alpha_\mathrm{MLT}=1.95$. Using this value, the dotted track in
Fig.~\ref{f:9} results. This would be the completely consistent,
correctly calculated track to be used in an update of globular cluster
age determinations. Obviously, the new opacity tables along with the
consistent $\alpha_\mathrm{MLT}$ is very close to the AF94-track, and
the effect on isochrones is minimal (e.g.\ effective temperatures
differ by 30~K at $\log L/L_sun = 2$), which is a very convenient, but
fortunate fact. Thus the globular ages determine in \citet{sw:98} and
\citet{sw:2002} are correct and remain unchanged. 
We also investigated how a recalibrated
$\alpha_\mathrm{MLT}$ would modify the effect on the more metal-rich
effect. This will be reported below and in Fig.~\ref{f:11}.

\begin{figure}
\begin{center}
\includegraphics[scale=0.65]{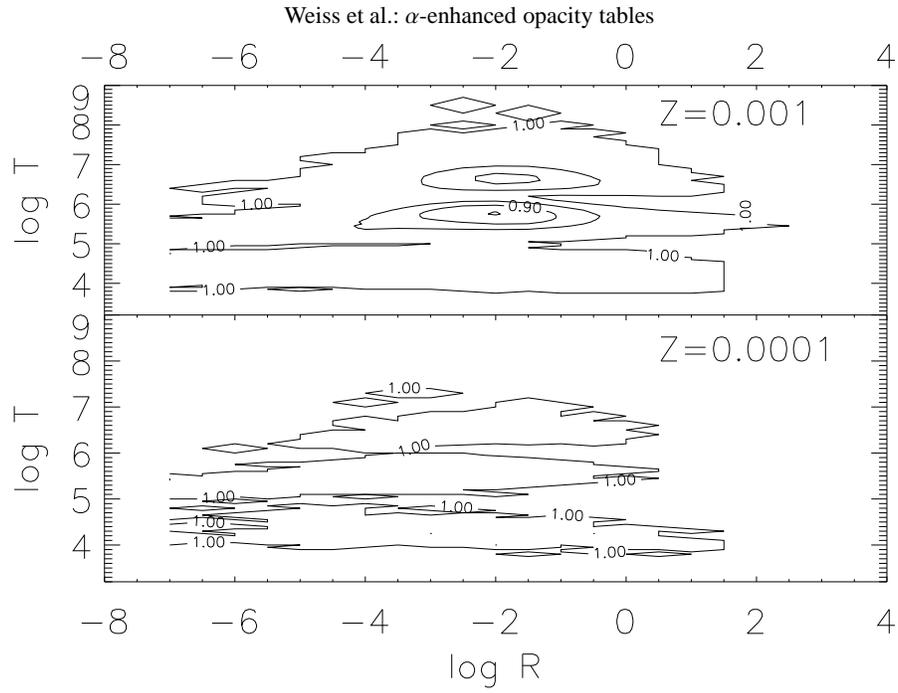}
\caption{As Fig.~\ref{f:4}, but for $Z=0.001$ and 0.0001. $X=0.70$ in
both panels.}
\label{f:10}
\end{center}
\end{figure}

The second aspect to be considered is that of the mixture effect in
the high-T opacities (Sect.~\ref{s:t2}), which could affect age
determinations of globular clusters by the increased main-sequence
lifetimes. The difference in the opacities, going from one
$\alpha$-enhanced mixture to the other, is vanishing with decreasing
metallicity (Fig.~\ref{f:10}). The difference for the lowest
metallicity considered here amounts to less than 3\% everywhere.
As a consequence, the turn-off ages agree within 200~Myr,
or about 2\% of the main-sequence lifetime.

The evolution shown in the upper panel of Fig.~\ref{f:9} demonstrates
the negligible effect of varying the internal element distribution on
the HRD of the $0.8\,\msun$ star: at fixed $Z=0.001$ the three tracks
using opacity tables for solar (dash-dotted line), $\alpha$-v
(long-dashed) and $\alpha$-c (dotted) element ratios are extremely close to each
other, the largest -- but still very small -- effect appearing at the turn-off: 
The effective temperature agrees for the scaled-solar and $\alpha$-c
mixture within 5~K, while the $\alpha$-v track is $\approx 30$K
hotter. Luminosities are higher by 0.005 resp.\ 0.014~dex.
This confirms again the results by \citet{ssc:93}.
In contrast, the
``generation effect'' of the AF94 track is much larger. No
recalibration of $\alpha_\mathrm{MLT}$ was done here; this would shift
all new tracks in the same way as the dotted one in the lower panel.

Given these results, we return briefly to the super-solar case. In
Fig.~\ref{f:11} we compare various cases for two mass values (0.8 and
$1.2\,\msun$). The solid line corresponds again to case B ($\alpha$-c mixture both in
the model and in the opacity tables). The fully equivalent models for
the $\alpha$-v mixture, taken again into account in both model and
opacities \citep{fa:2005} is indicated by the dotted line. In both
these cases $\alpha_\mathrm{MLT} = 1.6$. The recalibration of the
mixing length parameter yields a value close to 1.7; this case is
represented by the dashed line. Finally, case A (model composition
$\alpha$-c, old \citealt{af:94} tables,  $\alpha_\mathrm{MLT} = 1.6$)
is the most extreme case (dash-dotted line). In contrast to the
low-metallicity case, the recalibration of $\alpha_\mathrm{MLT}$ does
not approximate case A. It is also obvious that the change of mixture
has its largest effect on the main-sequence due to the increased
high-temperature opacities, but less so on along the RGB, where
temperatures are much more affected by $\alpha_\mathrm{MLT}$, which in
turn has a smaller effect along the main-sequence, in particular of
course for the higher mass with very thin exterior convection zone. 

\clearpage

\section{Discussion}

The diversity of chemical compositions for stellar models will remain
to be much larger than that of available opacity tables, which,
however are crucial for accurate model calculations. We have therefore
investigated how opacities calculated for similar, but not identical
$\alpha$-element enhanced mixtures influence the evolution of stellar
models of low mass and super-solar metallicity, where effects are
expected to be large. While the models had constant
$\alpha$-enhancement factors ($\alpha$-c; Table~\ref{t:1}) we used
tables for the same composition, but also for another one of varying
enhancement factors, $\alpha$-v. The latter opacity tables had been
computed about a decade ago with the \citet{af:94} code in response to
a specific request. 

The initial comparison (Fig.~\ref{f:1}) revealed drastic changes in
the temperature of red giants and the location and duration of the
main sequence. These changes were much larger than what could be
expected on the basis of the experience with using solar-scaled
mixture opacities instead of $\alpha$-enhanced ones
\citep{ssc:93,sw:98}. 

The origin of low- and high-temperature opacities differing, we
selectively replaced them to identify the cause for the differences
found. The lower stellar temperatures found in case of $\alpha$-c
opacity tables are, in particular along the RGB, almost exclusively
due to the low-T molecular opacities \citep{fa:2005}, while the main
sequence luminosity and thus lifetime varies with the composition of
the (OPAL-)high-T tables, the latter by up to 20\%! 
As the code computing the latter tables has not
changed since \cite{ir:96}, this was identified as a pure composition
effect: The $\alpha$-c opacities, for given $(X,Z)$, is consistently
higher in a temperature and density range covering the whole energy
producing core of low-mass main-sequence stars, which explains the
lower luminosity found. We have performed various tests to make sure
that we obtained and used the correct tables and have thus verified
that the {\em individual $\alpha$-element abundances play a
significant role for the opacities at stellar core temperatures}. 
In addition, we created opacity tables for the same mixtures using the
latest Opacity Project data \citep{bbb_op:2005,seaton:2005}. While in
the interesting temperature range the OP Rosseland mean opacities
differ from the OPAL ones, the {\em differential} effect going from
$\alpha$-v to $\alpha$-c abundance ratios is very similar. If
the opacities we have been using are confirmed to be true, the
conclusion is far-reaching: at solar- and super-solar metallicities,
the lifetime of $\alpha$-enriched low-mass stars is sensitive to the
individual abundance pattern. Accurate age determinations of stars or
stellar systems, such as elliptical galaxies or of the galactic bulge,
may require very detailed knowledge about the chemical composition.

For the low-T opacities we further had to discriminate between the
composition and the generation effect, as the $\alpha$-c tables had
been computed with the recent \citet{fa:2005} code. As the result of
our tests we discovered that the original $\alpha$-v tables suffer from an
error and are incorrect. Recomputations using both the new and the old
Wichita State molecular opacity code resulted in very similar
opacities, which, when used (Sects.~\ref{s:t3} and \ref{s:r}), yield much lower RGB
temperatures.\footnote{The old opacity tables for this mixture have
been removed from the web-site {\tt
http://webs.wichita.edu/physics/opacity/} and replaced by new calculations.}

The conclusion of our investigation is that at high metallicities
both the error in the old low-T tables and the composition details
in the high-T opacities significantly influence our models. Previous
calculations using the erroneous tables at solar metallicities should
be repeated with the new tables for the same $\alpha$-v
composition. For sub-solar metallicities both effects tend to
decrease, and in particular for typical Pop.~II compositions
variations in individual element abundances are no longer significant
for lifetimes. Age determinations of globular clusters are thus
independent of composition details. We therefore confirm once again
the basic results of \citet{ssc:93}. The erroneous low-T tables, in
contrast, still render the tracks too hot on the RGB, such that
colour-based age determinations would be affected (yielding ages too
low). Incidentally, however, a recalibration of the mixing length
parameter \amlt, which may be considered necessary when using the new
low-T tables of \citet{fa:2005} almost recovers the old RGB
temperatures (Fig.~\ref{f:9}, lower panel).

To summarize, we found that metal-rich low-mass stellar models are
very sensitive to details of the chemical composition used in opacity
tables. It is therefore advisable to employ  opacity tables consistent
with the assumed stellar chemical composition. The range of massive
stars still needs to be investigated. For population synthesis models
of metal-rich galaxies knowledge about the mean
$\alpha$-enrichment may not be sufficient for any reasonable
statements about the star formation history.

\begin{acknowledgements}
JF acknowledges support from NSF grant AST-0239590 with matching support
from the State of Kansas and the Wichita State University High Performance
Computing Center funded by NSF grants EIA-0216178 EPS-0236913.
\end{acknowledgements}
\bibliographystyle{aa}
\bibliography{weiss,cluster,oat}


\end{document}